# Impact of water on the charge transport of a glass-forming ionic liquid

P. Sippel[a,*], V. Dietrich[a], D. Reuter[a], M. Aumüller[a], P. Lunkenheimer[a], A. Loidl[a,b], S. Krohns [a,b]

[a] *Experimental Physics V, Center for Electronic Correlations and Magnetism, University of Augsburg, 86135 Augsburg, Germany*
[b] *Institute for Materials Resource Management, University of Augsburg, 86152 Augsburg, Germany*

Using dielectric spectroscopy and differential scanning calorimetry, we have performed a detailed investigation of the influence of water uptake on the translational and reorientational glassy dynamics in the typical ionic liquid 1-Butyl-3-methyl-imidazolium chloride. From a careful analysis of the measured dielectric permittivity and conductivity spectra, we find a significant acceleration of cation reorientation and a marked increase of the ionic conductivity for higher water contents. The latter effect mainly arises due to a strong impact of water content on the glass temperature, which for the well-dried material is found to be larger than any values reported in literature for this system. The fragility, characterizing the non-Arrhenius glassy dynamics of the ionic subsystem, also changes with varying water content. Decoupling of the ionic motion from the structural dynamics has to be considered to explain the results.

## 1. Introduction

In the past years, the investigation of ionic liquids has become one of the most popular fields in materials research [1,2]. The huge amount of publications on this topic points to the high scientific interest attracted by these materials and demonstrates their importance and versatility. Ionic liquids are molten salts at ambient temperatures. They have numerous potential applications, e.g., as solvents in green chemistry [3,4] or as electrolytes in electrochemical devices [5,6,7]. The latter one benefits from the low volatility and electrochemical stability of these solvent-free ionic conductors. Another key-factor for electrochemical applications is the ionic charge transport. Unfortunately, the conductivity observed in ionic liquids cannot compete with that of conventional electrolytes yet. However, there is a vast number of possible cations and anions that can be combined to form ionic liquids. This provides a high potential for designing ionic-liquid electrolytes that meet the benchmark for ionic conductivity, suitable for application [4,8,9].

Nearly all ionic liquids have the tendency to absorb water in moist atmosphere and thus water is considered to be ubiquitous even in ionic liquids regarded as hydrophobic [10,11,12,13]. Residual water is an important issue affecting their physical properties [11,13,14] like ionic mobility, which is significantly influenced even by small amounts [15]. For a successful technical application of ionic liquids, an eminent challenge is the careful control of water impurities [15,16]. So far, a lot of progress was made in determining the impact of water and organic solvents on viscosity and density of ionic liquids (e.g., refs. 11,13,17,18,19,20,21,22,23,24,25,26,27,28). In general, water impurities lead to a decrease of both quantities.

According to Sonnleitner *et al.* [29], dilution with a polar solvent "lubricates" an ionic liquid and its dynamics is getting considerably faster. Consequently, the conductivity should increase with higher water concentration, as was generally found in various earlier works, at least for a low impurity level [13,20,25,26,30,31,32,33,34,35]. According to Liu *et al.* [25], mixtures of ionic liquids and water display a maximum in the concentration dependence of the conductivity, similar to saline solutions. Maxima at higher water concentrations were also found in refs. [13] and [32]. Recently, we have reported a correlation between the conductivity and the glassy dynamics of aprotic ionic liquids [36]. Thus, varying the water content can be assumed to influence the glass temperature. Yaghini *et al.* [37] report a significant impact of water on the glass temperature in a protic ionic liquid. Wojnarowska *et al.* [38] found a strongly decreasing glass temperature when the amount of water increases in a lidocaine-based ionic liquid with hydrophilic chloride anions. Especially for these hydrophilic anions, the impact of water on the physical properties seems to be more pronounced [18].

Most ionic liquids show glassy freezing and dielectric spectroscopy is an ideal tool to probe their dynamic processes and conductivity. To our knowledge, until now only a few studies with dielectric spectroscopy have been performed on mixtures of ionic liquids and water and the majority of them investigates relatively high water concentrations [35,38,39,40,41]. In the present work, we focus on 1-Butyl-3-methyl-imidazolium chloride (BMIM Cl) and determine the impact of water uptake on the glass temperature, conductivity, and dipolar dynamics. Both liquids are freely miscible [18] and, based on the hygroscopic nature of the chloride anion, moisture from the surrounding atmosphere is absorbed up to a high concentration.

* Corresponding author.
E-mail address: pit.sippel@physik.uni-augsburg.de

## 2. Experimental details

The ionic liquid 1-Butyl-3-methyl-imidazolium chloride (BMIM Cl) was purchased from IoLiTec (Ionic Liquids Technologies GmbH, Heilbronn, Germany) with a purity of 99%. To vary the water content, the samples (about 3 ml, kept in a glass beaker with 1.5 cm diameter) were first dried in vacuum at 350 K at least for 24 hours and then exposed to air with approximately 40% relative humidity. Immediately before the performed differential scanning calorimetry (DSC) measurements, the water content of the samples was determined by coulometric Karl Fischer titration (KFT), executed with a C30 Compact KF Coulometer from Mettler Toledo. For comparison, one sample (termed "dry" in the following) was dried for another 24 hours at 420 K in vacuum to minimize the water concentration. However, due to the high viscosity of this sample, no KFT measurement could be performed.

To check for glass and possible phase transitions, the samples were characterized by DSC between 323 and 103 K with a scan rate of 10 K/min. The samples were sealed in aluminum pans and a power compensating DSC 8500 (Perkin Elmer) was used. Moreover, the kinetic fragility was determined for three samples by the Wang-Velikov fictive-temperature method [42]. For this purpose, the cooling rate was varied between 0.5 K/min and 100 K/min and a constant heating rate of 10 K/min was used.

The spectra of the complex permittivity and of the conductivity were measured between 1 Hz and 10 MHz using a frequency-response analyzer (Novocontrol Alpha-analyzer). Dielectric spectroscopy at additional frequencies up to 200 MHz was performed for sample 1 with an Agilent 4991A impedance analyzer based on the I-V technique, with the sample mounted at the end of a specially designed coaxial line [43]. In all cases, the sample material was filled into parallel-plate capacitors. For cooling and heating, a nitrogen gas cryostat (Novocontrol Quatro) was used.

## 3. Results and discussion

### 3.1. DSC and KFT

BMIM Cl was reported to be easily supercooled and to transform into a solid glass at low temperatures [44]. In agreement with ref. [44], the present DSC measurements performed with 10 K/min only revealed a glass transition but no melting or freezing points. This was observed for all investigated samples with various water contents, which was determined by KFT. Fig. 1(a) shows a magnified view of the DSC results for different mixtures of BMIM Cl and water (72 mol%, 57 mol%, 9 mol%, 1 mol% and the "dry" sample) in the vicinity of their glass transitions. The curves where obtained by heating the sample with 10 K/min after cooling from ambient temperature with the same rate. All samples show a well-pronounced step-like anomaly in the heat flow associated with the glass transition. The glass temperatures $T_g$, defined by the onset of the anomaly, are indicated by arrows. For the "dry" sample (dashed line) the highest glass temperature ($T_g = 269$ K) is found. With increasing water content, the glass temperature of BMIM Cl significantly shifts to lower temperatures. The water absorption of the ionic liquid in air (with approximately 40% relative humidity) saturates at 72 mol%, where the glass transition has dropped to 172 K (solid line). It should be noted that the glass temperatures of 197 - 208 K, reported for BMIM Cl in earlier works, are clearly lower than $T_g = 269$ K found for the dry sample in the present work [18,44,45,46]. As denoted, e.g., in ref. [36], $T_g$ has a direct impact on the conductivity of ionic liquids at room temperature. Thus, the strong concentration dependence of the glass temperature as documented in Fig. 1(a) demonstrates the importance of knowing and documenting the water concentration of hygroscopic ionic liquids, which is often neglected in literature [5].

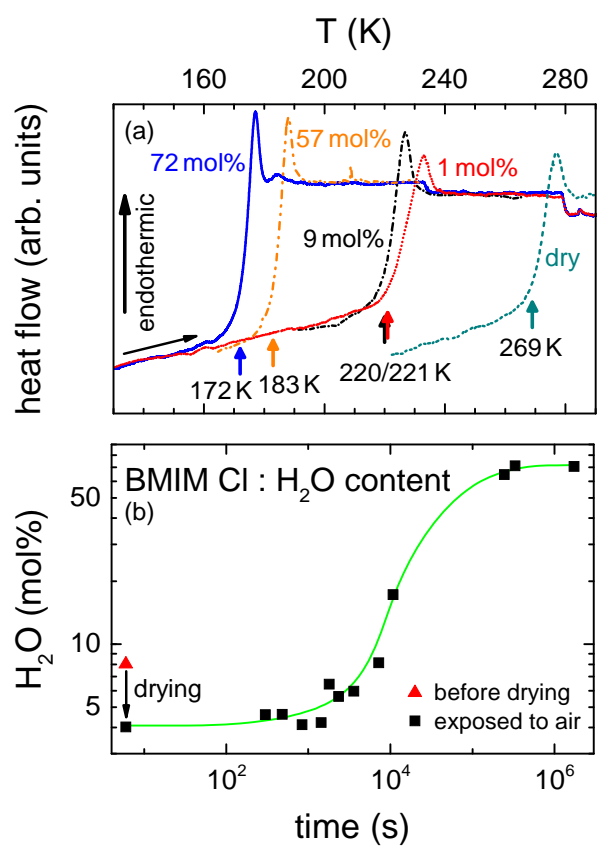

**Fig. 1.** (a) DSC heating traces revealing the glass transitions for BMIM Cl with various water contents (the high viscosity of the "dry" sample did not allow for determination of the water content). Arrows mark the onset of glass transitions. (b) Time-dependent water absorption in air with approximately 40 % relative humidity at room temperature determined by KFT (line to guide the eye). The triangle shows the initial water content of the sample material before drying at 350 K.

With increasing water content, $T_g$ reaches 172 K, dropping by more than 35% if compared to the "dry" sample. To our knowledge, until now a similar impact of water on the glass-transition temperature was only reported for the protic ionic liquid lidocaine hydrochloride [38], which also contains a chloride anion. A water-induced reduction of the glass-transition temperature may be typical for ionic liquids, at least for this type of anion. However, for less hydrophilic anions even the opposite $T_g$ shift was observed [18] and, obviously, the influence of water on the glass transition temperatures in ionic liquids is a complex issue. In general, the viscosity of ionic liquids decreases with increasing water content [17,22,23,24,27]. However, the temperature dependence of their viscosity changes as well [22], which could affect the glass temperature and may explain these opposing trends.

The sample of BMIM Cl, taken directly from the desiccator without any additional drying procedure, contains 8 mol% water (Fig. 1(b), triangle). This demonstrates that applying a proper drying procedure is prerequisite for meaningful measurements of pure ionic liquids. To estimate the timescale for water absorption, a dried sample was humidified by moist air (see section 2) and the water concentration was probed at different times via KFT. The time-dependent water content of BMIM Cl is displayed in a double-logarithmic representation in Fig. 1(b). Initial drying was performed at 350 K for 24 hours in reduced pressure (transition from triangle to square, marked by an arrow) finally leading to water content of about 4 mol%. Within the first 10 minutes the water content turned out to be almost constant. One reason may be the relatively high viscosity of the dried sample, hampering water diffusion within the sample. After more than 60 minutes the water content increased significantly as indicated by the s-shaped line in Fig. 1(b). The original water concentration before drying was recovered after subjecting the sample for approximately 2 hours to this atmosphere. After more than 90 hours, the water content saturated at about 70 mol%. Notably, handling the sample several minutes in air does not drastically change the initial $H_2O$ concentration. However, even small changes in water content can significantly affect the sample properties. This is revealed, e.g., by the DSC measurements shown in Fig. 1(a) when comparing the additionally dried (24 hours at 420 K) and 1 mol% sample.

*3.2. Temperature-dependent dielectric properties*

Fig. 2 shows the temperature dependence of the dielectric constant and conductivity of BMIM Cl with water (sample 2) for various frequencies. This sample was taken right out the desiccator without any further treatment. It is supposed to have an initial water content of about 8 mol% (see Fig. 1(b)); nevertheless the exact concentration during the dielectric measurement is not known as it could have changed significantly during preparation and while being exposed to the nitrogen atmosphere in the cryostat prior to and during the measurement. As, however, we found the curves from the performed cooling and heating runs matching each other (at least up to 270 K; in Fig. 2 only the cooling curve is shown), the water concentration at least did not dramatically change during the measurement. At temperatures below about 180 K, $\varepsilon'$ is almost temperature and frequency independent (Fig. 2(a)) with a value of approximately 3, representing the high-frequency limit of the dielectric constant, $\varepsilon_\infty$, arising from the electronic and ionic polarizability. With increasing temperature, the permittivity shows a step-like rise and even exceeds values of $10^6$ for the 1 Hz curve. This feature shifts to higher temperatures with increasing frequency, thus revealing the typical signature of a relaxation process. The detected high values of the dielectric constant are of non-intrinsic nature and caused by the effect of electrode polarization (blocking electrodes) as found in many ion conductors [47,48].

A closer look at $\varepsilon'(T)$ shown in Fig. 2(a) reveals that the observed increase is most likely composed of several successive steps that superimpose each other. This is indicated, e.g., by the change of slope located around 250 K for the 1 Hz curve. Indications for another relaxational contribution are found at high frequencies, e.g., below 260 K for 67 kHz: The shallow variation of $\varepsilon'(\nu)$ at the base of the main relaxation feature seems to signify a further relaxation step with a saturation value $\varepsilon_s$ of the order of 10. This superimposed process is best revealed by a detailed equivalent-circuit analysis of the frequency-dependent data as will be discussed in section 3.3.

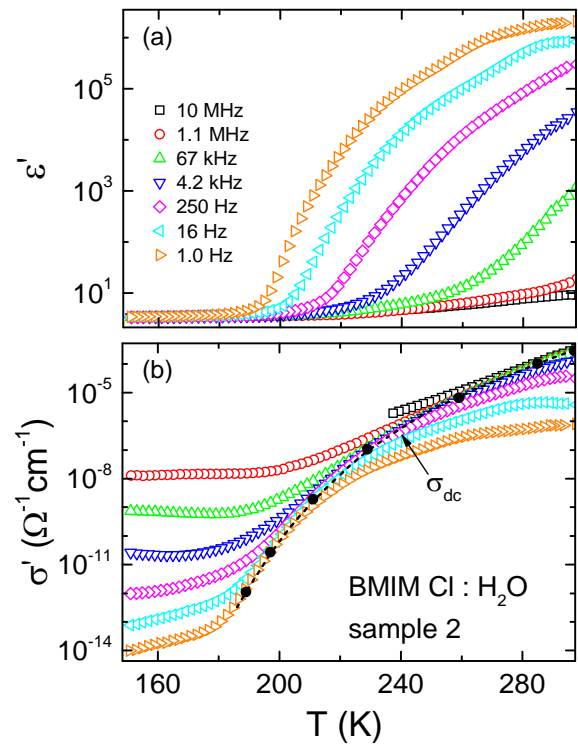

**Fig. 2.** Temperature dependence of the dielectric constant (a) and conductivity (b) of BMIM Cl with water (untreated sample 2) for various frequencies. The dashed line represents an educated guess of the dc conductivity. The closed circles show the dc conductivity as obtained from the fits of the frequency-dependent data (Fig. 3).

The presence of pronounced electrode-polarization effects, which lead to so-called colossal values in $\varepsilon'$ [49], is typical for ionic liquids and superimposes the intrinsic sample properties. For this reason, the rather controversially debated modulus representation is sometimes used to characterize the ionic dynamics [50,51,52,53,54]. In this representation the contributions from electrode polarization are suppressed. However, in the present report we prefer using the conductivity representation instead as it allows for a more direct determination of the technically relevant dc-conductivity, $\sigma_{dc}$, showing up as a frequency-independent region in plots of $\sigma'(T,\nu)$.

The temperature-dependent conductivity is shown in Fig. 2(b). At low temperatures, plateau-like regions show up, whose absolute values strongly increase with increasing frequency. Such a strong frequency dependence of the conductivity at low temperatures and high frequencies could arise from the so-called "universal dielectric response" (UDR) [55], i.e., a power law $\sigma'(\nu) \propto \nu^s$ with $s < 1$, which is often caused by ionic hopping conductivity [56,57,58]. However, in Fig. 2(b) weak, peak-like structures in $\sigma'(T)$ are superimposed to the plateaus (e.g., at about 170 K for 1.1 MHz), which shift to lower temperatures with decreasing frequency. Thus an intrinsic relaxational contribution seems to play a role here, which is confirmed by the analysis of the frequency-dependent plots as discussed in section 3.3. At higher temperatures, $T > 190$ K, the plateaus are followed by frequency-independent regions of the conductivity, i.e., the same temperature dependence of $\sigma'(T)$ is observed for different frequencies (e.g., for 1 Hz and 16 Hz around 200 K). The corresponding frequency-independent temperature profile represents the dc conductivity $\sigma_{dc}(T)$ of the sample and is indicated by the dashed line in Fig. 2(b). At temperatures above 200 K, $\sigma'(T)$ at the lowest frequencies becomes smaller than $\sigma_{dc}$. This reduction of the conductivity at low frequencies is due to the onset of electrode polarization, which leads to colossal values of $\varepsilon'$ (Fig. 2(a)) as discussed above [48].

Finally, it should be noted that for a sample of 8% water content, the glass temperature should be of the order of 220 K (Fig. 1(a)). Usually, below $T_g$ all temperature-dependent quantities exhibit a crossover to weaker temperature dependence, which can be ascribed to the sample falling out of equilibrium and fictive temperature starting to deviate from actual temperature when further cooling below $T_g$. However, no anomaly in $\varepsilon'(T)$ or $\sigma'(T)$ is observed around 220 K in Fig. 2, indicating a significantly lower glass temperature, probably due to an uptake of water during preparation as mentioned in the beginning of this section.

### 3.3 Frequency-dependent dielectric properties

Fig. 3 shows the spectra at 1 Hz - 10 MHz of the real (a) and imaginary part (b) of the dielectric permittivity ($\varepsilon'$ and $\varepsilon''$, respectively) and the conductivity $\sigma'$ (c) as obtained for sample 2 at selected temperatures between 297 and 157 K. While, in principle, the information contained in $\varepsilon''$ and $\sigma'$ is redundant (because $\sigma' \propto \varepsilon'' \nu$), the separate discussion of both is helpful because the various dielectrically active processes occurring in ionic liquids are differently emphasized in these quantities.

The huge increase of $\varepsilon'(\nu)$ towards lower frequencies observed in Fig. 3(a) is caused by blocking electrodes, as mentioned in section 3.2. They give rise to a so-called Maxwell-Wagner (MW) relaxation which, in principle, should lead to relaxation steps in $\varepsilon'(\nu)$ with a well-defined plateau at low frequencies [59]. However, even at the highest temperatures investigated, the spectra in Fig. 3(a) do not approach a well-defined plateau, a phenomenon that is often observed for ionically conducting materials [48]. This effect can be ascribed to a distribution of MW-relaxation times as discussed, e.g., in ref. [48]. Corresponding to this smeared-out step in $\varepsilon'$, instead of a distinct local maximum a broad shoulder is observed in $\varepsilon''(\nu)$ (Fig. 3(b)) (e.g., at about 1 kHz for the 285 K curve).

The high-frequency flanks of these shoulders in $\varepsilon''(\nu)$ arise from the dc conductivity, which becomes more obvious in the conductivity representation of Fig. 3(c). Here, well-defined frequency-independent plateaus shows up, corresponding to $\sigma_{dc}$. Notably, due to blocking-electrode effects leading to a drop of $\sigma'(\nu)$ at low frequencies, for ionic conductors usually ac measurements are necessary to access the dc conductivity [36]. Indeed, for high temperatures, $T \geq 211$ K, the conductivity shown in Fig. 3(c) is strongly decreasing at low frequencies, as the electrode polarization leads to an insulating layer in series to the bulk sample. In the dielectric loss in Fig. 3(b), this feature corresponds to the levelling-off of $\varepsilon''(\nu)$ at low frequencies and high temperatures. As already suggested based on the temperature-dependent plot of $\sigma' \propto \varepsilon'' \nu$ (Fig. 2(b)), the frequency-dependent representation of $\varepsilon''$ provides evidence for an intrinsic relaxational process evolving at the lowest temperatures. It shows up in $\varepsilon''$ at 157 K as a clear maximum at $\nu > 100$ kHz.

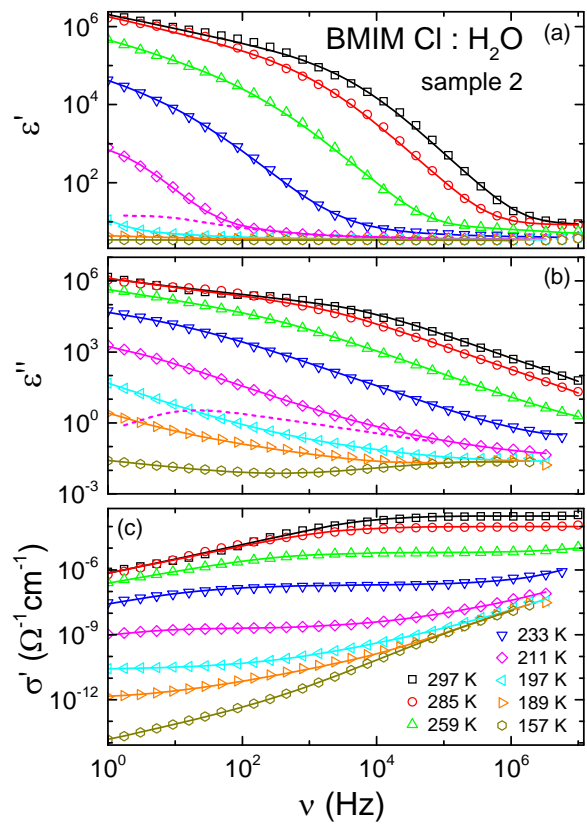

**FIG. 3.** Spectra of the dielectric constant (a), loss (b), and conductivity (c) of BMIM Cl with water (untreated sample 2), measured at various temperatures. The solid lines are fits assuming a distributed RC circuit in series with the sum of a CD and a CC function and a contribution from dc charge transport [48,59]. The dashed lines in frames (a) and (b) indicate the contribution from the $\alpha$ relaxation for 211 K.

For a detailed analysis, the spectra of Fig. 3 were simultaneously fitted in $\varepsilon'(\nu)$ and $\varepsilon''(\nu)$ using an equivalent circuit [49,59]. To account for the dominating MW relaxation, we employed a distributed RC circuit [48,60], connected in series to the bulk sample. The bulk properties were modeled by a dc resistor and two relaxational functions. One of the latter accounts for the mentioned relaxation, clearly revealed at low temperatures and high frequencies. In ref. [45], this process was discussed in terms of a secondary Johari-Goldstein relaxation [61], which is found in nearly all types of glassy matter [45,62,63]. Alternatively, an intramolecular mode of the cation was considered. To fit this relaxation, we used a Cole-Cole (CC) function [64], which is known to provide a good description of secondary relaxations [63]. The second relaxation at lower frequencies with $\varepsilon_s \approx 10$, already mentioned in the discussion of the temperature-dependent data, is assumed to arise from the reorientational motion of dipolar cations [45,65]. In correspondence to the nomenclature used for dipolar supercooled liquids it will be termed $\alpha$ relaxation in the following. However, it should be noted that in this ionic liquid, it only mirrors one aspect of the glassy dynamics, namely the reorientation of the cations. For the present analysis, it was modeled by a Cole-Davidson (CD) function [66] as often employed for $\alpha$ processes in glassy matter [67,68]. This relaxation process leads, e.g., to a smearing out of the onset of the MW relaxation steps in $\varepsilon'(\nu)$ (Fig. 3(a)). Its contributions in $\varepsilon'(\nu)$ and $\varepsilon''(\nu)$ are indicated for 211 K by the dashed lines in Figs. 3(a) and (b). Due to the superposition by the dominating MW relaxation (in $\varepsilon'$) and by the dc conductivity (in $\varepsilon''$ and $\sigma'$), the presence of this relaxation process is not immediately obvious in the spectra of Fig. 3. However, fits without this second relaxation contribution did not lead to a satisfactory agreement with the experimental data.

The fit curves obtained by this model are in good agreement with the experimental data as demonstrated by solid lines in Fig. 3. Overall, the found dielectric properties are typical for ionic liquids and a similar succession of dynamic processes was reported in various works [36,45,65,69,70,71,72,73,74,75]. For imidazolium-based ionic liquids, additional modes arising from librations and low-energy 'intermolecular' vibrations have been found [70,71,72] to occur at high frequencies, $\nu > 1$ GHz, which, however, are out of the scope of the present work.

It should be noted that the experimental spectra could be fitted without any UDR contribution (see section 3.2). Obviously, the nearly temperature-independent, strongly frequency-dependent regions in $\sigma'(T)$ revealed by Fig. 2(b) are completely explainable by the secondary relaxation, which dominates in the corresponding frequency/temperature range. For the lowest temperature of 157 K, it corresponds to the region of about 1 kHz to 3 MHz in the $\sigma'(\nu)$ plot of Fig. 3(c), exhibiting a nearly linear increase due to the strong broadening of the secondary relaxation peak.

*3.4 Water-dependent dielectric properties*

Fig. 4 shows the dielectric spectra of three samples of BMIM Cl with varying water content as measured at three selected temperatures. As mentioned above, the initial water content of sample 2, after taking it out of the desiccator, should have been about 8 mol% (see Fig. 1(b)) but the actual concentration may be higher. To ensure the lowest water content, sample 1 was dried inside the nitrogen gas cryostat by keeping it at 373 K for 20 hours. Between drying and the subsequent measurements, the sample was not removed from the cryostat to avoid uncontrolled water absorption prior to the dielectric measurement. Sample 3 was exposed to air (~ 40% relative humidity) until 12 mol% water content was reached. Sample preparation was performed in the same atmosphere, immediately before the dielectric measurement. Sample 3 thus should have the highest water content. All measurements were performed in nitrogen atmosphere, both under cooling and heating. The temperature-dependent curves from both runs were found to agree well for sample 1 and 2, demonstrating that the water concentration did not significantly change during the measurements. For sample 3, however, the cooling and heating curves did not completely agree, pointing to a minor reduction of water content in the dry nitrogen atmosphere. In Fig. 4, we show data collected during cooling.

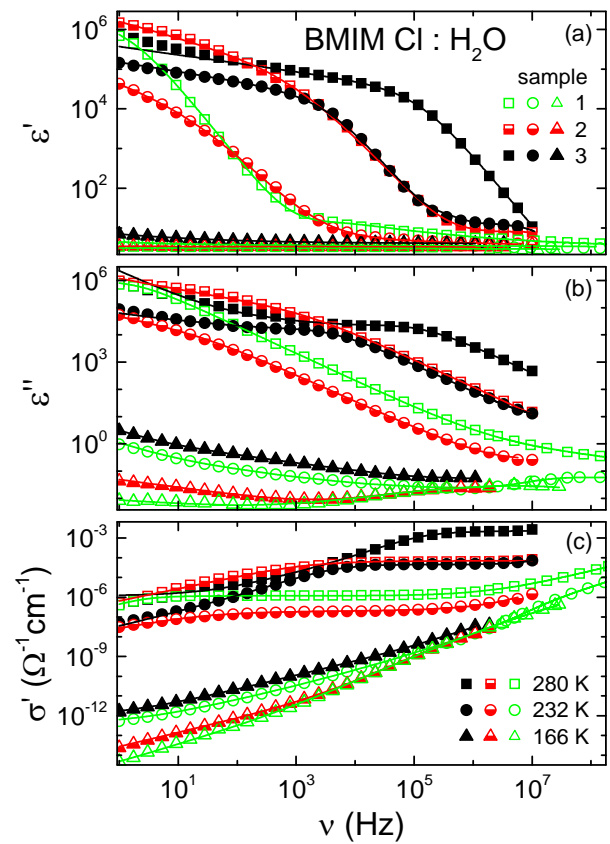

**Fig. 4.** Spectra of the dielectric constant (a), loss (b), and conductivity (c) of BMIM Cl with different water contents for selected temperatures (sample 1 has the lowest and sample 3 the highest water content). The lines are fits with a distributed RC circuit in series with the sum of a CD and a CC function or two CC functions and a dc-conductivity contribution [48,59].

The squares in Fig. 4(a) show $\varepsilon'(\nu)$ for the three samples at 280 K. At this temperature it becomes obvious that the MW relaxation arising from electrode polarization strongly shifts towards lower frequencies with decreasing water content. This is also revealed in $\varepsilon''(\nu)$ and $\sigma'(\nu)$ (Figs. 4(b) and (c)), where the accompanying peaks or shoulders are shifted by about five decades in frequency when comparing sample 1 and 3. Taking into account that the viscosity of ionic liquids raises with reducing water content [17,22,23,24,27], this behavior could be expected: For an increased viscosity, the translational motion of the ions should become slower and, thus, electrode polarization occurs at lower frequencies. A closer inspection of Fig. 4 reveals differences in the spectral shapes of the electrode effects for the three samples. Since electrode polarization is of non-intrinsic origin, it is highly sensitive to the surface of the electrodes and geometry and, thus, the corresponding features in the spectra can alter in different measurements.

Just as the MW relaxation, the reorientational mode of the cations also slows down with decreasing water content. This trend is, e.g., revealed by the 280 K curves in Fig. 4(a): For sample 3 the corresponding relaxation step with $\varepsilon_s \approx 10$ obviously is located at frequencies above 10 MHz while its low-frequency plateau just becomes visible for sample 2 (above about 1 MHz). Finally, for sample 1, the step is revealed at about 3 kHz - 10 MHz. According to the Debye-Stokes-Einstein relation, the rotational relaxation time of a particle in a liquid medium can be assumed to be proportional to viscosity and particle volume. While deviations from this relationship in ionic liquids are known [76], it at least is approximately valid as explicitly shown for different alkyl-chain lengths of imidazolium-based ionic liquids [65,77]. Thus, the observed shift of the $\alpha$ relaxation can be related to the mentioned water-induced viscosity variation. When assuming the approximate validity of the Stokes-Einstein relation, connecting viscosity and translational diffusion coefficient, the marked reduction of the dc plateaus with decreasing water content, documented in Fig. 4(c), can be explained, too: For higher viscosity the translational mobility of the ionic charge carriers is reduced and the conductivity becomes smaller.

The spectra of all samples are well described by fits as discussed in section 3.3, assuming a distributed RC circuit for the blocking electrodes [48], dc conductivity, and two or three relaxational functions (lines in Fig. 4). In some cases, a third relaxation process located between the $\alpha$ relaxation and the mentioned secondary relaxation process was necessary to account for the experimental data. Interestingly, in ref. [45] a second secondary process was also found for BMIM-BMSF (BMSF stands for bis(trifluoromethane sulphonate)imide) and ascribed to an intrinsic Johari-Goldstein process arising from the presence of a nonsymmetric anion. However, this explanation cannot be applied to the present case with a spherical Cl$^-$ anion. Obviously, more work is necessary to clarify the presence and microscopic origins of secondary relaxations in these systems, which, however, is out of the scope of the present work.

As the main focus of this work is the investigation of the influence of water on the conductivity, here we only briefly discuss the reorientational relaxation times $\tau$ as obtained from the fits. Fig. 5 shows the deduced average relaxation-time values $\langle\tau\rangle$ in an Arrhenius representation. ($\langle\tau\rangle = \tau$ for the CC function and $\langle\tau\rangle = \beta\tau$ for the CD function, where $\beta$ is the broadening parameter.) The figure also presents $\tau_\alpha$ values for the $\alpha$ relaxation, arising from the main reorientational cation motion, and $\tau_s$ for the faster, well pronounced secondary relaxation. The relaxation times of the mentioned additional third process, which is strongly superimposed by other contributions, have very high uncertainty and thus are not included in Fig. 5. In accord with the above discussion of Fig. 4, the $\alpha$-relaxation times (closed symbols in Fig. 5) are found to be highly sensitive to the different water content of the samples. For all three samples, the temperature dependence of $\tau_\alpha(T)$ can be well fitted by the empirical Vogel-Fulcher-Tammann law (solid lines) [78,79,80,81],

$$\tau = \tau_0 \exp\left[\frac{DT_{VF}}{T - T_{VF}}\right], \qquad (1)$$

which is commonly employed to parameterize the typical non-Arrhenius temperature dependence of glassy dynamics. Here $\tau_0$ corresponds to an inverse attempt frequency [82], $T_{VF}$ is a divergence temperature, and $D$ is the so-called strength parameter introduced by Angell [81]. The latter is used to distinguish between strong and fragile glass formers and is a measure for the deviation from Arrhenius behavior [81]. Another way to parameterize these deviations is the fragility index $m$, which is defined as the slope at $T_g$ in the Angell plot, log $\tau$ vs. $T_g/T$ [83]. Higher values of $m$ or lower values of $D$ correspond to more fragile temperature characteristics and stronger deviations from Arrhenius behavior.

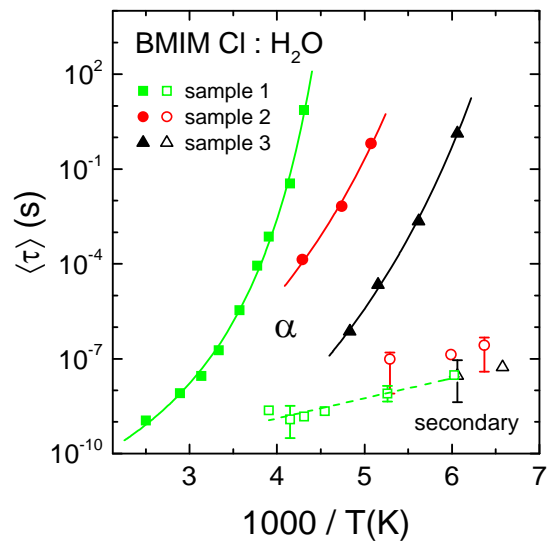

**Fig. 5.** Average reorientational relaxation times of BMIM Cl with various water contents in an Arrhenius representation as obtained from the fits of the dielectric spectra (sample 1 has the lowest and sample 3 the highest water content). The $\alpha$-relaxation times $\tau_\alpha$ (closed symbols) are fitted by VFT laws (solid lines). The open symbols are the relaxation times $\tau_s$ of the high-frequency secondary relaxation. The dashed line is a linear fit of the data for sample 1, corresponding to an energy barrier of 0.13 eV.

We find a clear decrease of $T_{VF}$ for increasing water content (sample 1: 181 K, sample 2: 118 K, sample 3: 105 K). The strength parameter increases ($D$ = 8.3, 18, and 20, respectively) but for samples 2 and 3 its absolute value is of limited significance due to the small number of data points. The secondary relaxation times (open symbols in Fig. 5) exhibit much weaker temperature dependence than $\tau_\alpha$ and approximately follow an Arrhenius law. Due to the superposition with the stronger $\alpha$ relaxation in the spectra, they have rather large uncertainty. Taking into account the corresponding error bars shown in Fig. 5, no significant variation of $\tau_s$ with water content could be detected.

*3.5. Water-dependent conductivity and glassy dynamics*

Among the quantities determined from the fits of the spectra, $\sigma_{dc}$ is the most interesting one since it is of technical relevance for many applications of ionic liquids. Fig. 6 shows the temperature dependence of the dc conductivity in an Arrhenius representation as obtained from the fits of the spectra for samples 1 - 3. For sample 2, the same data points are also shown in Fig. 2(b) (closed circles), demonstrating that the fit results reasonably agree with the estimate of $\sigma_{dc}(T)$ based on the temperature-dependent plot of $\sigma'$ (dashed line). In Fig. 6, all samples reveal a non-linear behavior as also reported for many other ionic liquids [36,52,54,71,74,84,85,86,87,88,89,90]. It can be described by the empirical Vogel-Fulcher-Tammann (VFT) formula, written in the modified form:

$$\sigma_{dc} = \sigma_0 \exp\left[\frac{-DT_{VF}}{T - T_{VF}}\right] \quad (2)$$

In Fig. 6, the VFT fits are extrapolated to $\sigma_{dc} \approx 10^{-15}$ $\Omega^{-1}$cm$^{-1}$ (the start value of the ordinate in Fig. 6). For ionic conductors where the ionic motion is well coupled to the structural relaxation, $\sigma_{dc}$ is expected to approach this value at $T_g$ [91]. For systems with decoupling of structural and ionic dynamics, the temperatures obtained by this extrapolation may be regarded as an estimate of the glass-transition temperature of the ionic subsystem [36,92]. We obtain 229 K (sample 1), 176 K (sample 2), and 159 K (sample 3). This variation is consistent with the strong decrease of $T_g$ with increasing water content found in the DSC experiments (Fig. 1). Possible decoupling effects will be discussed below.

Fig. 6 clearly reveals that sample 1 (with the lowest water content) has the lowest dc conductivity in the whole temperature range. At high temperatures, $\sigma_{dc}$ of the moistened sample 2 is about one order of magnitude higher. With decreasing temperature, the difference between $\sigma_{dc}$ of these two samples strongly increases and at 250 K it exceeds five decades. The dc conductivity of sample 3 with the highest water content is even higher. Samples 2 and 3 seem to exhibit a more moderate temperature dependence of $\sigma_{dc}$ than sample 1. The technically relevant room-temperature conductivity increases almost by a factor of 500, namely from $1.5\times10^{-5}$, over $3.4\times10^{-4}$, to $7.0\times10^{-3}$ $\Omega^{-1}$cm$^{-1}$ for samples 1, 2, and 3, respectively. This is mainly due to the mentioned marked reduction of $T_g$ with increasing water content.

The glass temperatures of samples 1 and 2 are within the $T_g$ range deduced from the DSC results indicated by the double arrow in Fig. 6. As mentioned in section 3.2, for sample 2, which was measured without further treatment after taking it out of the desiccator, an initial water concentration of about 8 mol% is expected. Thus, its glass temperature should be of the order of 220 K (Fig. 1(a)), i.e., much higher than $T_g \approx 176$ K, deduced from Fig. 6. This marked discrepancy certainly is partly due to some water uptake that occurred during sample preparation for the dielectric measurements as discussed in section 3.3. However, the results on sample 3 indicate that additional effects may play a role: The glass temperature of 159 K deduced from the conductivity for this sample (Fig. 6) is far out of the range of the DSC experiments, where even for the saturated sample, with 72% water content, a higher $T_g$ of 172 K was found. Therefore it is unreasonable to assume that sample 3 should have even higher water content. Obviously, there must be an additional effect that causes the deviations of the glass temperatures determined from the conductivity and DSC. As mentioned above, the conductivity extrapolation in principle can only reveal $T_g$ of the ionic subsystem. Therefore, a solution for this discrepancy is a decoupling of the translational ionic motion from the structural relaxation that is sensed by the DSC measurements. Similar effects have been observed for many ionic conductors [91,93,94,95,96,97], including various ionic liquids [38,52,54,76,85,86,98,99]. When scaled onto each other, often the conductivity exhibits a weaker temperature dependence than the viscosity and/or structural relaxation time, when approaching the glass transition. Therefore the decoupling results in lower glass temperatures for the conductivity as indeed found in the present work. However, it should be mentioned that Wojnarowska *et al.* [38] did not find any decoupling between conductivity and structural relaxation times for various water contents in a lidocaine-based ionic liquid.

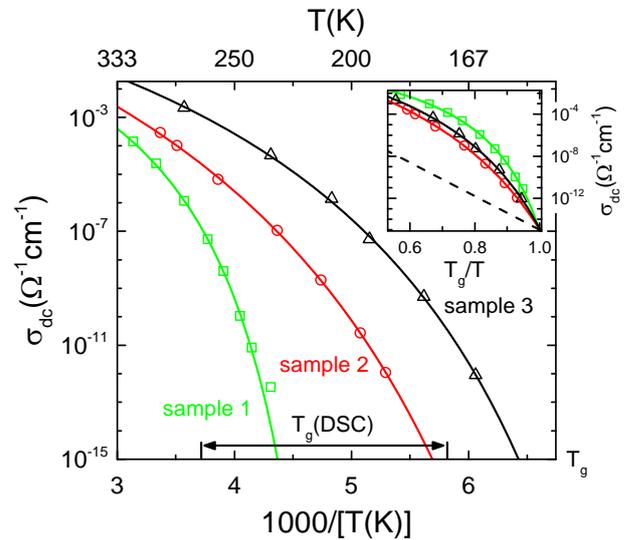

**Fig. 6.** Temperature-dependent dc-conductivity of BMIM Cl with various water contents in an Arrhenius representation (sample 1 has the lowest and sample 3 the highest water content). The lines are fits using the VFT equation, Eq. (2). The start value of the ordinate ($10^{-15}$ $\Omega^{-1}$cm$^{-1}$) corresponds to the conductivity at the glass transition for the fully coupled case [91]. The double arrow marks the range of glass-transition temperatures obtained by DSC (Fig. 1(a)). The inset is an Angell plot of the same data with the dashed line representing a glass former with maximum strength ($m = 16$).

Together with the glass temperature, the fragility is an essential quantity for the room-temperature conductivity of ionic liquids [36]. Using the $T_g$ values determined from Fig. 6, the Angell plot for $\sigma_{dc}$ shown in the inset of Fig. 6 reveals a variation of fragility for the three samples: Within the experimental resolution it is similar for samples 2 and 3 ($m \approx 50\pm5$) but for sample 1, with the lowest water content, it is clearly higher ($m \approx 80\pm5$). To rationalize this finding, we speculate that in samples 2 and 3, containing relatively large amounts of water, an extended hydrogen-bonded network of water molecules may form. The rather well-defined short-range order in such hydrogen-bonded networks leads to a lower density of energy minima in configuration space than in the nearly pure ionic liquid (sample 1), where isotropic ionic interactions dominate. According to ref. [100], this should result in a reduction of fragility as observed for the samples with high water content. As demonstrated in ref. [36], for ionic liquids with similar glass temperature, a higher fragility can significantly reduce the room-temperature dc-conductivity. In the present case, however, it is obvious, that the conductivity enhancement of the moistened samples is dominated by their strongly reduced glass temperature.

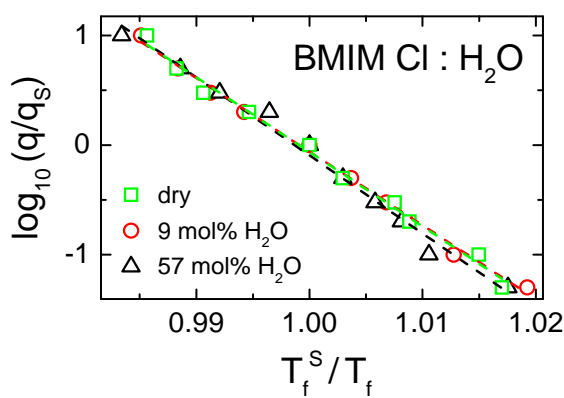

**Fig. 7.** Arrhenius plot of the reduced cooling rate *vs.* the reduced fictive temperature for BMIM Cl with varying water content. The scaling parameters are the standard cooling rate $q_s = 10$ K/min and the corresponding standard fictive temperature $T_f^s$. The lines are linear fits, corresponding to fragilities of $m = 68\pm2$ (dry), $67\pm2$ (9 mol%), and $71\pm3$ (57 mol%).

It should be noted that the fragility discussed in the preceding paragraph was deduced from the conductivity and thus is a property of the ionic subsystem. To obtain more insight into the fragility associated with the structural relaxation of these mixtures, we performed temperature-rate dependent DSC experiments, which enable determining the fragility around the glass temperature (details on the procedure can be found in literature [42,101,102,103]). In ref. [86] this method was applied to ionic liquids. Fig. 7 illustrates the results in an Arrhenius plot of the reduced cooling rate *vs.* the reduced fictive temperature. In this representation, the fragility index can be obtained from the slope of a linear fit. DSC curves at the standard rate of 10 K/min of these samples are displayed in Fig. 1(a). Interestingly, Fig. 7 reveals no significant change in the fragility index for the different samples ($m$ only varies between 68 and 71), although their water concentration is drastically different and the fragility obtained from the dc conductivity is changing in this regime. This finding again points to a decoupling of structural and ionic dynamics in these systems.

## 4. Summary

A detailed investigation of the dielectric and conductivity properties of the prototypical ionic liquid BMIM Cl with varying water contents has revealed a strong impact of water on the glassy dynamics and ionic charge transport. An equivalent-circuit analysis of the dielectric spectra, including a proper modelling of the observed blocking-electrode effects, enabled the deconvolution of the contributions from different dynamics processes. The most dramatic water-induced effects obviously arise from the strong reduction of the glass temperature, dropping from 269 to 172 K under water uptake from air. The result for the best dried sample exceeds the various $T_g$ values reported in literature for this system. The observed strong enhancement of ionic conductivity by several decades is mainly due to this variation of $T_g$. We also found a water-induced change of fragility of the ionic subsystem. The fact that this is not reproduced by the fragility determined from cooling-rate dependent DSC experiments indicates decoupling of the ionic and structural dynamics. This is corroborated by a comparison of $T_g$ deduced from the DSC and the conductivity experiments. The reorientational motion of the cations is also strongly affected by the water content and becomes faster by several decades. Overall, the results of the present work indicate a dramatic impact of water uptake on the glassy and ionic dynamics, demonstrating the importance of a proper characterization of water content in ionic liquids.

**Acknowledgments**

We thank M. Weiss for performing parts of the dielectric measurements. This work was supported by the BMBF via ENREKON 03EK3015.

**References**

[1] F. Endres, S.Z.E. Abedin, Air and water stable ionic liquids in physical chemistry, Phys. Chem. Chem. Phys. 8 (2006) 2101-2116.
[2] R.D. Rogers, S.J. Zhang, J.J. Wang, Preface: An international look at ionic liquids, Sci. China Chem. 55 (2012) 1475-1477.
[3] R.D. Rogers, K.R. Seddon, Ionic liquids - Solvents of the future?, Science 302 (2003) 792-793.
[4] H. Weingärtner, Understanding ionic liquids at the molecular level: facts, problems, and controversies, Angew. Chem. Int. Ed. 47 (2008) 654-670.
[5] M. Armand, F. Endres, D.R. MacFarlane, H. Ohno, B. Scrosati, Ionic-liquid materials for the electrochemical challenges of the future, Nature Mater. 8 (2009) 621-629.
[6] D.R. MacFarlane, N. Tachikawa, M. Forsyth, J.M. Pringle, P.C. Howlett, G.D. Elliott, J.H. Davis, M. Watanabe, P. Simon, C.A. Angell, Energy applications of ionic liquids, Energy Environ. Sci. 7 (2014) 232-250.
[7] M.-C. Lin, M. Gong, B.G. Lu, Y.P. Wu, D.-Y. Wang, M.Y. Guan, M. Angell, C.X. Chen, J. Yang, B.-J. Hwang, H.J. Dai, An ultrafast rechargeable aluminium-ion battery, Nature 520 (2015) 325-328.
[8] P. Simon, Y. Gogotsi, Materials for electrochemical capacitors, Nature Mater. 7 (2008) 845–854.

[9] C. Chiappe, D. Pieraccini, Ionic liquids: solvent properties and organic reactivity, J. Phys. Org. Chem. 18 (2005) 275-297.
[10] L. Cammarata, S.G. Kazarian, P.A. Salter, T. Welton, Molecular states of water in room temperature ionic liquids, Phys. Chem. Chem. Phys. 3 (2001) 5192-5200.
[11] J.A. Widegren, A. Laesecke, J.W. Magee, The effect of dissolved water on the viscosities of hydrophobic room-temperature ionic liquids, Chem. Commun. (2005) 1610-1612.
[12] T. Köddermann, C. Wertz, A. Heintz, R. Ludwig, The association of water in ionic liquids: A reliable measure of polarity, Angew. Chem. Int. ed. 45 (2006) 3697-3702.
[13] E.P. Grishina, L.M. Ramenskaya, M.S. Gruzdev, O.V. Kraeva, Water effect on physicochemical properties of 1-butyl-3-methylimidazolium based ionic liquids with inorganic anions, J. Mol. Liq. 177 (2013) 267-272.
[14] J.L. Anthony, E.J. Maginn, J.F. Brennecke, Solution thermodynamics of imidazolium-based ionic liquids and water, J. Phys. Chem. B 105 (2001) 10942-10949.
[15] Y. Danten, M.I. Cabaco, M. Besnard, Interaction of water diluted in 1-butyl-3-methyl imidazolium ionic liquids by vibrational spectroscopy modeling, J. Mol. Liq. 153 (2010) 57-66.
[16] J.E.S.J. Reid, A.J. Walker, S. Shimizu, Residual water in ionic liquids: clustered or dissociated?, Phys. Chem. Chem. Phys. 17 (2015) 14710-14718.
[17] K.R. Seddon, A. Stark, M.J. Torres, Influence of chloride, water and organic solvents on the physical properties of ionic liquids, Pure Appl. Chem. 72 (2000) 2275-2287.
[18] J.G. Huddleston, A.E. Visser, W.M. Reichert, H.D. Willauer, G.A. Broker, R.D. Rogers, Characterization and comparison of hydrophilic and hydrophobic room temperature ionic liquids incorporating the imidazolium cation, Green Chem. 3 (2001) 156-164.
[19] J.J. Wang, Y. Tian, Y. Zhao, K. Zhuo, A volumetric and viscosity study for the mixtures of 1-n-butyl-3-methylimidazolium tetrafluoroborate ionic liquid with acetonitrile, dichloromethane, 2-butanone and N,N-dimethylformamide, Green Chem. 5 (2003) 618-622.
[20] H.T. Xu, D.C. Zhao, P. Xu, F.Q. Liu, G. Gao, Conductivity and viscosity of 1-allyl-3-methyl-imidazolium chloride plus water and plus ethanol from 293.15 K to 333.15 K, J. Chem. Eng. Data, 50 (2005) 133-135.
[21] N. Calvar, B. Gonzalez, A. Dominguez, J. Tojo, Physical properties of the ternary mixture ethanol+water+1-butyl-3-methylimidazolium chloride at 298.15 K, J. Solution Chem. 35 (2006) 1217-1225.
[22] H. Rodriguez, J.F. Brennecke, Temperature and composition dependence of the density and viscosity of binary mixtures of water plus ionic liquid, J. Chem. Eng. Data 51 (2006) 2145-2155.
[23] J. Jacquemin, P. Husson, A.A.H. Padua, V. Majer, Density and viscosity of several pure and water-saturated ionic liquids, Green Chem. 8 (2006) 172-180.
[24] C. Schröder, T. Rudas, G. Neumayr, S. Benkner, O. Steinhauser, On the collective network of ionic liquid/water mixtures. I. Orientational structure, J. Chem. Phys. 127 (2007) 234503.
[25] W.W. Liu, L.Y. Cheng, Y.M. Zhang, H.P. Wang, M.F. Yu, The physical properties of aqueous solution of room-temperature ionic liquids based on imidazolium: Database and evaluation, J. Mol. Liq. 140 (2008) 68-72.
[26] J.N.C. Lopes, M.F.C. Gomes, P. Husson, A.A.H. Padua, L.P.N. Rebelo, S. Sarraute, M. Tariq, Polarity, viscosity, and ionic conductivity of liquid mixtures containing [C(4)C(1)im][Ntf(2)] and a molecular component, J. Phys. Chem. B 115 (2011) 6088-6099.
[27] A.L. Sturlaugson, K.S. Fruchey, M.D. Fayer. Orientational dynamics of room temperature ionic liquid/water mixtures: Water-induced structure. J. Phys. Chem. B 116, 1777-1787 (2012).
[28] U. Domańska, M. Królikowska, Density and viscosity of binary mixtures of thiocyanate ionic liquids plus water as a function of temperature, J. Solution Chem. 41 (2012) 1422-1445.
[29] T. Sonnleitner, D.A. Turton, S. Waselikowski, J. Hunger, A. Stoppa, M. Walther, K. Wynne, R. Buchner, Dynamics of RTILs: A comparative dielectric and OKE study, J. Mol. Liq. 192 (2014) 19-25.
[30] S.D. Mikhailenko, S. Kaliaguine, J.B. Moffat, Electrical impedance studies of the ammonium salt of 12-tungstophosphoric acid in the presence of liquid water, Solid State Ion. 99 (1997) 281-286.
[31] J. Bowers, C.P. Butts, P.J. Martin, M.C. Vergara-Gutierrez, R.K. Heenan, Aggregation behavior of aqueous solutions of ionic liquids, Langmuir 20 (2004) 2191-2198.
[32] A. Jarosik, S.R. Krajewski, A. Lewandowski, P. Radzimski. Conductivity of ionic liquids in mixtures, J. Mol. Liq. 123 (2006) 43-50.
[33] J.J. Wang, H.Y. Wang, S.L. Zhang, H.H. Zhang, Y. Zhao, Conductivities, volumes, fluorescence, and aggregation behavior of ionic liquids [C$_4$mim][BF$_4$] and [C$_n$mim]Br ($n$ = 4, 6, 8, 10, 12) in aqueous solutions, J. Phys. Chem. B 111 (2007) 6181-6188.
[34] H. Shekaari, S.S. Mousavi, Y. Mansoori, Thermophysical properties of ionic liquid, 1-pentyl-3-methylimidazolium chloride in water at different temperatures, Int. J. Thermophys. 30 (2009) 499-514.
[35] A. Swiety-Pospiech, Z. Wojnarowska, S. Hensel-Bielowka, J. Pionteck, M. Paluch, Effect of pressure on decoupling of ionic conductivity from structural relaxation in hydrated protic ionic liquid, lidocaine HCl, J. Chem. Phys. 138 (2013) 204502.
[36] P. Sippel, P. Lunkenheimer, S. Krohns, E. Thoms, A. Loidl, Importance of liquid fragility for energy applications of ionic liquids, Sci. Rep. 5 (2015) 13922.
[37] N. Yaghini, J. Pitawala, A. Matic, A. Martinelli, Effect of water on the local structure and phase behavior of imidazolium-based protic ionic liquids, J. Phys. Chem. B 119 (2015) 1611-1622.
[38] Z. Wojnarowska, K. Grzybowska, L. Hawelek, A. Swiety-Pospiech, E. Masiewicz, M. Paluch, W. Sawicki, A. Chmielewska, P. Bujak, J. Markowski, Molecular dynamics studies on the water mixtures of pharmaceutically important ionic liquid lidocaine HCl, Mol. Pharm. 9 (2012) 1250-1261.
[39] M. Bester-Rogac, A. Stoppa, J. Hunger, G. Hefter, R. Buchner, Association of ionic liquids in solution: a combined dielectric and conductivity study of [bmim][Cl] in water and in acetonitrile, Phys. Chem. Chem. Phys. 13 (2011) 17588-17598.
[40] X.Q. Fan, K.S. Zhao, Thermodynamics of micellization of ionic liquids C$_6$mimBr and orientation dynamics of water for C$_6$mimBr-water mixtures: A dielectric spectroscopy study, J. Phys. Chem. B 118 (2014) 13729-13736.
[41] X.Q. Fan, K.S. Zhao, Aggregation behavior and electrical properties of amphiphilic pyrrole-tailed ionic liquids in water, from the viewpoint of dielectric relaxation spectroscopy, Soft Matter 10 (2014) 3259-3270.
[42] L.-M. Wang, V. Velikov, C.A. Angell, Direct determination of kinetic fragility indices of glassforming liquids by differential scanning calorimetry: Kinetic versus thermodynamic fragilties, J. Chem. Phys. 117 (2002) 10184-10192.
[43] R. Böhmer, M. Maglione, P. Lunkenheimer, A. Loidl, Radio-frequency dielectric measurements at temperatures from 10 to 450 K, J. Appl. Phys. 65 (1989) 901-904.
[44] C.P. Fredlake, J.M. Crosthwaite, D.G. Hert, S.N.V.K. Aki, J.F. Brennecke, Thermophysical properties of imidazolium-based ionic liquids, J. Chem. Eng. Data 49 (2004) 954-964.
[45] A. Rivera, E.A. Rössler, Evidence of secondary relaxations in the dielectric spectra of ionic liquids, Phys. Rev. B 73 (2006) 212201.
[46] U. Domańska, E. Bogel-Lukasik, R. Bogel-Lukasik, 1-octanol/water partition coefficients of 1-alkyl-3-methylimidazolium chloride, Chem. Eur. J. 9 (2003) 3033-3041.
[47] J.R. Macdonald, Comparison and discussion of some theories of the equilibrium double layer in liquid electrolytes, J. Electroanal. Chem. 223 (1987) 1-23.
[48] S. Emmert, M. Wolf, R. Gulich, S. Krohns, S. Kastner, P. Lunkenheimer, A. Loidl, Electrode polarization effects in broad band dielectric spectroscopy, Eur. Phys. J. B 83 (2011) 157-165.
[49] P. Lunkenheimer, V. Bobnar, A.V. Pronin, A.I. Ritus, A.A. Volkov, A. Loidl, Origin of apparent colossal dielectric constants, Phys. Rev. B 66 (2002) 052105.
[50] P.B. Macedo, C.T. Moynihan, R. Bose, The role of ionic diffusion in polarisation in vitreous ionic conductors, Phys. Chem. Glasses 13 (1972) 171-179.
[51] I.M. Hodge, K.L. Ngai, C.T. Moynihan, Comments on the electric modulus function, J. Non-Cryst. Solids 351 (2005) 104.
[52] A. Rivera, A. Brodin, A. Pugachev, E.A. Rössler, Orientational and translational dynamics in room temperature ionic liquids, J. Chem. Phys. 126 (2007) 114503.
[53] J.R. Sangoro, F. Kremer, Charge transport and glassy dynamics in ionic liquids, Acc. Chem. Res. 45 (2011) 525-531.
[54] Z. Wojnarowska, M. Paluch, Recent progress on dielectric properties of protic ionic liquids, J. Phys.: Condens. Matter 27 (2015) 073202.
[55] A. K. Jonscher, Universal dielectric response, Nature (London) 267 (1977) 673-679.
[56] S.R. Elliott, A.P. Owens, The diffusion-controlled relaxation model for ionic transport in glasses, Phil. Mag. B 60 (1989) 777-792.
[57] K. Funke, Ion transport and relaxation studied by high-frequency conductivity and quasi-elastic neutron scattering, Phil. Mag. A 64 (1991) 1025-1034.

[58] J.C. Dyre, P. Maass, B. Roling, D.L. Sidebottom, Fundamental questions relating to ion conduction in disordered solids, Rep. Prog. Phys. 72 (2009) 046501.
[59] P. Lunkenheimer, S. Krohns, S. Riegg, S.G. Ebbinghaus, A. Reller, A. Loidl, Colossal dielectric constants in transition-metal oxides, Eur. Phys. J. Special Topics 180 (2010) 61-89.
[60] P. Lunkenheimer, G. Knebel, A. Pimenov, G.A. Emelchenko, A. Loidl, Dc and ac conductivity of $La_2CuO_{4+\delta}$, Z. Phys. B: Condens. Matter 99 (1996) 507-516.
[61] G. P. Johari, M. Goldstein, Viscous liquids and the glass transition. II. Secondary relaxations in glasses of rigid molecules, J. Chem. Phys. 53 (1970) 2372-2388.
[62] A. Kudlik, S. Benkhof, T. Blochowicz, C. Tschirwitz, E. Rössler, The dielectric response of simple organic glass formers, J. Mol. Struct. 479 (1999) 201-218.
[63] S. Kastner, M. Köhler, Y. Goncharov, P. Lunkenheimer, A. Loidl, High-frequency dynamics of type-B glass formers investigated by broadband dielectric spectroscopy, J. Non-Cryst. Solids 357 (2011) 510.
[64] K.S. Cole, R.H. Cole, Dispersion and absorption in dielectrics I. Alternating current characteristics, J. Chem. Phys. 9 (1941) 341-351.
[65] K. Nakamura and T. Shikata, Systematic dielectric and NMR study of the ionic liquid 1-alkyl-3-methyl imidazolium, ChemPhysChem 11 (2010) 285-294.
[66] D.W. Davidson, R.H. Cole, Dielectric relaxation in glycerol, propylene glycol, and n-propanol, J. Chem. Phys. 19 (1951) 1484-1490.
[67] P. Lunkenheimer, U. Schneider, R. Brand and A. Loidl, Glassy dynamics, Contemp. Phys. 41 (2000) 15-36.
[68] F. Kremer, A. Schönhals, Analysis of dielectric spectra, in: F. Kremer, A. Schönhals (Eds.), Broadband dielectric spectroscopy, Springer, Berlin, 2009 Ch. 3, pp. 59-98.
[69] H. Weingärtner, A. Knocks, W. Schrader, U. Kaatze, Dielectric spectroscopy of the room temperature molten salt ethylammonium nitrate, J. Phys. Chem A 105 (2001) 8646-8650.
[70] A. Stoppa, J. Hunger, R. Buchner, G. Hefter, A. Thoman, H. Helm, Interactions and dynamics in ionic liquids, J. Phys. Chem. B 112 (2008) 4854-4858.
[71] J. Sangoro, C. Iacob, A. Serghei, S. Naumov, P. Galvosas, J. Kärger, C. Wespe, F. Bordusa, A. Stoppa, J. Hunger, R. Buchner, F. Kremer, Electrical conductivity and translational diffusion in the 1-butyl-3-methylimidazolium tetrafluoroborate ionic liquid, J. Chem. Phys. 128 (2008) 214509.
[72] R. Buchner, G. Hefter, Interactions and dynamics in electrolyte solutions by dielectric spectroscopy, Phys. Chem. Chem. Phys. 11 (2009) 8984-8999.
[73] J. Hunger, A. Stoppa, S. Schrödle, G. Hefter, R. Buchner, Temperature dependence of the dielectric properties and dynamics of ionic liquids. ChemPhysChem 10 (2009) 723-733.
[74] C. Krause, J.R. Sangoro, C. Iacob, F. Kremer, Charge transport and dipolar relaxations in imidazolium-based ionic liquids, J. Phys. Chem. B 114 (2010) 382–386.
[75] H. Weingärtner, The static dielectric permittivity of ionic liquids, J. Mol. Liq. 192 (2014) 185-190.
[76] N. Ito, R. Richert, Solvation dynamics and electric field relaxation in an imidazolium-$PF_6$ ionic liquid: from room temperature to the glass transition, J. Phys. Chem. B 111 (2007) 5016-5022.
[77] C. Daguenet, P.J. Dyson, I. Krossing, A. Oleinikova, J. Slattery, C. Wakai, H. Weingärtner, Dielectric response of imidazolium-based room-temperature ionic liquids, J. Phys. Chem. B 110 (2006) 12682-12688.
[78] H. Vogel, The law of the relationship between viscosity of liquids and the temperature, Phys. Z. 22 (1921) 645-646.
[79] G.S. Fulcher, Analysis of recent measurements of the viscosity of glasses, J. Am. Ceram. Soc. 8 (1925) 339-355.
[80] G. Tammann, W. Hesse, Die Abhängigkeit der Viscosität von der Temperatur bei unterkühlten Flüssigkeiten, Z. Anorg. Allg. Chem. 156 (1926) 245-257.
[81] C.A. Angell, Strong and fragile liquids, in: K.L. Ngai and G.B. Wright (Eds.), Relaxations in complex systems, NRL, Washington DC, 1985, pp. 3-11.
[82] To account for the small number of data points, for sample 2 $\tau_0$ was fixed to $8 \times 10^{-13}$ s.
[83] R. Böhmer, K.L. Ngai, C.A. Angell, D.J. Plazek, Nonexponential relaxations in strong and fragile glass formers, J. Chem. Phys. 99 (1993) 4201-4209.
[84] C.A. Angell, Y. Ansari, Z. Zhao, Ionic Liquids: Past, present and future, Faraday Discuss. 154 (2012) 9-27.
[85] W. Xu, E.I. Cooper, C.A. Angell, Ionic liquids: Ion mobilities, glass temperatures, and fragilities, J. Phys. Chem. B 107 (2003) 6170-6178.
[86] K. Ueno, T. Zhao, M. Watanabe, C.A. Angell, Protic ionic liquids based on decahydroisoquinoline: Lost superfragility and ionicity-fragility correlation, Phys. Chem. B 116 (2012) 63-70.
[87] J. Sangoro, A. Serghei, S. Naumov, P. Galvosas, J. Kärger, C. Wespe, F. Bordusa, F. Kremer, Charge transport and mass transport in imidazolium-based ionic liquids, Phys. Rev. E 77 (2008) 051202.
[88] J.R. Sangoro, C. Iacob, A. Serghei, C. Friedrich, F. Kremer. Universal scaling of charge transport in glass-forming ionic liquids, Phys. Chem. Chem. Phys. 11 (2008) 913-916.
[89] J. Leys, M. Wübbenhorst, C.P. Menon, R. Rajesh, J. Thoen, C. Glorieux, P. Nockemann, B. Thijs, K. Binnemans, S. Longuemart, Temperature dependence of the electrical conductivity of imidazolium ionic liquids, J. Chem. Phys. 128 (2008) 064509.
[90] J. Leys, R.N. Rajesh, P.C. Menon, C. Glorieux, S. Longuemart, P. Nockemann, M. Pellens, K. Binnemans, Influence of the anion on the electrical conductivity and glass formation of 1-butyl-3-methylimidazolium ionic liquids, J. Chem. Phys. 133 (2010) 034503.
[91] F. Mizuno, J.-P. Belieres, N. Kuwata, A. Pradel, M. Ribes, C.A. Angell, Highly decoupled ionic and protonic solid electrolyte systems, in relation to other relaxing systems and their energy landscapes, J. Non-Cryst. Solids 352 (2006) 5147-5155.
[92] C.A. Angell, The nature of glassforming liquids, the origin of superionics and 'tight' vs. 'loose' glassy conductors, Solid State Ionics **105** (1998) 15.
[93] C.T. Moynihan, N. Balitactac, L. Boone, T.A. Litovitz, Comparison of shear and conductivity relaxation times for concentrated lithium chloride solutions, J. Chem. Phys. 55 (1971) 3013-3019.
[94] C.A. Angell, Fast ion motion in glassy and amorphous materials, Solid State Ionics 9-10 (1983) 3-16.
[95] C.A. Angell, Mobile ions in amorphous solids, Annu. Rev. Phys. Chem. 43 (1992) 693-717.
[96] G. Tarjus, D. Kivelson, Breakdown of the Stokes-Einstein relation in supercooled liquids, J. Chem. Phys. 103 (1995) 3071-3073.
[97] A. Pimenov, P. Lunkenheimer, H. Rall, R Kohlhaas, A. Loidl, R. Böhmer, Ion transport in the fragile glass-former $3KNO_3-2Ca(NO_3)_2$, Phys. Rev. E 54 (1996) 676-684.
[98] P.J. Griffin, A.L. Agapov, A.P. Sokolov, Translation-rotation decoupling and nonexponentiality in room temperature ionic liquids, Phys. Rev. E 86 (2012) 021508.
[99] Z. Wojnarowska, K. Kołodziejczyk, K.J. Paluch, L. Tajber, K. Grzybowska, K.L. Ngai, M. Paluch. Decoupling of conductivity relaxation from structural relaxation in protic ionic liquids and general properties, Phys. Chem. Chem. Phys. 15 (2013) 9205-9211.
[100] C.A. Angell, Perspective on the glass transition, J. Phys. Chem. Solids 49 (1988) 861-871.
[101] V. Velikov, S. Borick, C.A. Angell. The glass transition of water, based on hyperquenching experiments, Science 294 (2001) 2335-2338.
[102] V. Velikov, S. Borick, C.A. Angell. Molecular glasses with high fictive temperatures for energy landscape evaluations, J. Phys. Chem. B 106 (2002) 1069-1080.
[103] Y.Z. Yue, J.deC. Christiansen, S.L. Jensen, Determination of the fictive temperature for a hyperquenched glass, Chem. Phys. Lett. 357 (2002) 20-24.